\documentclass[prl,twocolumn,showpacs,preprintnumbers,amsmath,amssymb,superscriptaddress]{revtex4}
\usepackage{bm}
\usepackage{epsfig,amsopn}
\usepackage{graphicx}
\usepackage{amsmath,amssymb}
\usepackage{natbib}
\renewcommand{\section}[1]{{\par\it #1.---}}
\def\be{\begin{equation}}
\def\ee{\end{equation}}
\def\bea{\begin{eqnarray}}
\def\eea{\end{eqnarray}}
\def\la{\langle}
\def\ra{\rangle}
\def\om{\omega}
\def\nn{\nonumber}

\def\f{\frac}
\def\g{\gamma}
\def\al{\alpha}
\def\etal{{\emph{et al~}}}

\begin{document}

\title{ Heat conduction in the disordered Fermi-Pasta-Ulam
  chain}
\author{Abhishek Dhar}
\email{dabhi@rri.res.in}
\affiliation{Raman Research Institute, Bangalore 560080}

\author{Keiji Saito}
\affiliation{Graduate School of Science, 
University of Tokyo, 113-0033, Japan} 
\affiliation{CREST, Japan Science and Technology (JST), Saitama, 332-0012, Japan}
\date{\today} 
\begin{abstract}
We address the question of the effect of disorder on heat conduction
in an anharmonic chain with interactions given by the Fermi-Pasta-Ulam
(FPU) potential. In contrast to the conclusions  of an earlier paper
[{\bf   Phys. Rev. Lett.  86,  63 (2001)}]   which
found that disorder could induce a finite thermal conductivity at low temperatures, 
we find no evidence of a finite temperature  transition in conducting
properties. Instead, we find that at low temperatures,  small
system size transport properties are dominated by disorder but the asymptotic system 
size dependence of current is  given by the usual FPU result $J
\sim 1/N^{2/3}$. We also present new interesting results on the
binary-mass ordered FPU chain.
\end{abstract}

\maketitle
It is now generally believed that heat conduction in one-dimensional
($1D$) systems is anomalous \cite{LLP03,grass02}. In the absence of an external pinning potential, 
as is the case 
in most realistic situations, one finds that Fourier's law is not
 satisfied. One of  
the predictions from Fourier's law is the scaling form of heat current $J$ 
 with system size $N$ for a system with a fixed applied temperature difference. 
From Fourier's law one gets $J \sim 1/N$ . The conclusion, 
from a large number of studies of $1D$ momentum conserving systems seems to be:
\bea
J \sim \f{1}{N^{1-\alpha}} ~~~~~\alpha \neq 0~.
\eea
The main results on the exponent $\alpha$ can be summarized 
as follows: 
(1) for a pure harmonic chain $J \sim N^0$ \cite{rieder67}, (2) for a
disordered harmonic 
chain  $\alpha$ depends on the spectral  properties of the bath and on
 boundary conditions \cite{rubin71,casher71,dhar01}, and (3) for
 a nonlinear system without disorder 
 $\alpha$  seems to be independent of properties of heat baths and the results
 from the most recent simulations indicate a universal value of
 $\al=1/3$ \cite{grass02,mai07}. 

Transport in systems with both disorder and interactions has recently
attracted a lot of interest both theoretically
\cite{baowenli01,maynard90,pikovsky07,dhar08,basko06} and experimentally
\cite{baenninger08}. The  
main interest is to understand the transition from an insulating state
governed by the physics of Anderson localization to a conducting state
as one increases interactions.  
In the context of oscillator chains we note that the physics of the
disordered harmonic chain is dominated by 
localization physics which has its strongest effect in $1D$ systems.
The question of the effect of anharmonicity  on localization was recently 
addressed for a system where the harmonic part of the Hamiltonian
included an external  pinning potential  and the anharmonicity was a
quartic onsite potential \cite{dhar08}.
In this case, in the absence of the anharmonic term,  $J \sim e^{-c N}$. 
Surprisingly it was found that  adding a small amount of anharmonicity leads to 
 a conducting (Fourier-like) behaviour with a power
law decay $J \sim 1/N$ and no transition to the insulating state was
found on decreasing the anharmonicity. An important feature seen was that with   
decreasing anharmonicity one had  to go to larger system sizes to see the 
true asymptotic behaviour of the current. 

In the present letter we investigate the same question, namely that of
the effect of  
interactions (phonon-phonon) on localization, but in the absence of
any external pinning potential.
We study the mass disordered FPU model with interactions 
put in through a quartic inter-particle potential.
In the absence of pinning, for the harmonic case, low frequency modes with $\om
\stackrel{<}{\sim} 1/N^{1/2}$ remain
extended and this gives rise to a power law dependence of $J$ on $N$
\cite{rubin71,casher71,dhar01}.
On the other hand, for the pure FPU chain also, low frequency modes
are believed to play an important  role in transport and give rise to
anomalous transport.  
An earlier study by Li {\emph{et al}} \cite{baowenli01} on this model concluded that this model
showed a  transition, from a 
Fourier like scaling $J \sim 1/N$ at low temperatures, to a pure FPU
like behaviour  with $J \sim 1/N^{0.57}$ at high temperatures.
Our study suggests that this conclusion may not be correct. We do
not find any evidence of a finite temperature transition. Instead we
find that a small amount of   
anharmonicity leads to  the same system size dependence as seen in the pure
system. We discuss  possible sources of error in the conclusions
of Li {\emph{et al}}. We also present new results on the ordered
binary mass FPU chain including nontrivial scaling properties of the
system size dependence of current. 

\section{Model} We consider the following FPU Hamiltonian:
\bea
H \! &=&\! \sum_{l=1}^{N} \f{p_l^2}{2 m_l} 
+ \sum_{l=1}^{N+1} 
[ \f{(x_l-x_{l-1})^2}{2} +  \nu \f{(x_l-x_{l-1})^4}{4} ],~~~ 
\eea
where $\{ x_l, p_l \}$ denotes the position and momenta of particles
and we use fixed boundary conditions 
$x_0=x_{N+1}=0$. The interparticle harmonic spring constant has been set to one
and $\nu$ denotes the strength of the quartic interaction. We consider
a binary random alloy and set the masses of half of the particles, at
randomly chosen sites, to $m_1$ and the rest to $m_2$.
The particles at the two ends of the chain are 
connected to stochastic white noise heat baths at different temperatures. 
The equations of motion of the chain are then given by:
\bea
\label{eq:2}
m_l \ddot{x}_l&=&- (2 x_l-x_{l-1}-x_{l+1}) \nn \\
&-& \nu [ (x_l-x_{l-1})^3 + (x_l-x_{l+1})^3 ]-\g_l \dot{x}_l 
+\eta_l~,  ~~ \label{langevin}
\eea
with $\eta_l=\eta_L \delta_{l,1}+\eta_R \delta_{l,N},~\g_l=\g (\delta_{l,1}+
\delta_{l,N})$, 
 and where the noise terms satisfy the  fluctuation dissipation 
relations $\la \eta_L(t) \eta_L(t') \ra = 2 \g k_B T_L \delta(t-t')$,
$\la \eta_R(t) \eta_R(t') \ra = 2 \g k_B T_R \delta(t-t')$, $k_B$
being Boltzmann's constant. The heat current  is given by
$ J = \sum_l \la f_{l,l-1} \dot{x}_l \ra/(N-1)$ where $f_{l,l-1}$
is the force exerted by the $(l-1)$th particle on the $l$th particle
and $\la ...\ra$ denotes a steady state average. We will denote by
$[J]$ an average over disorder. 
As noted in \cite{dhar08}, Eqs.~(\ref{langevin}) are invariant under
the transformation $T_{L,R} \to s T_{L,R}$, 
$\{ x_l\} \to \{ s^{1/2} x_l \}$ and $\nu \to \nu/s$.  This implies the
scaling relation $J(sT_L,sT_R,\nu)= s J(T_L,T_R,s\nu)$ and thus the
effect of changing $\nu$ can be equivalently  studied by changing
$T_L,T_R$. In the present study we will fix temperatures and consider
the effect of changing $\nu$. 

Let us consider first the  harmonic case $\nu=0$.
In this case it is known from detailed numerical work and analytic
arguments that the exponent $\alpha$ depends on the properties of the
bath and on boundary conditions. For white noise baths one finds
$\alpha=-1/2$ for fixed boundaries and $\alpha=1/2$ for
free boundaries. In the presence of anharmonicity it is expected, and
indeed we have verified in simulations, that $\alpha$ does not
depend on boundary conditions. Here we use fixed boundary conditions only. 

\begin{figure}
\vspace{1cm}
\includegraphics[width=3in]{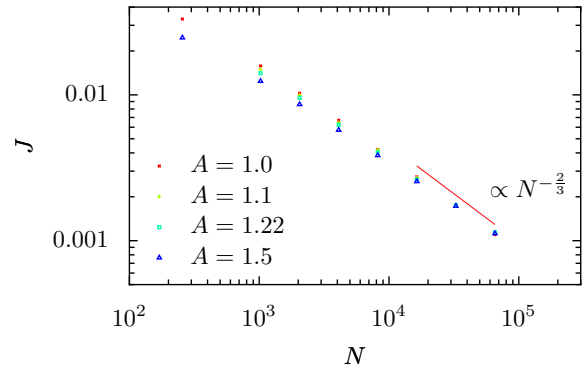}
\caption{Plot of the heat current $J$ versus
system size in the binary mass ordered chain for different values of
the mass ratio $A=1.0,1.1,1.22$ and $1.5$. }
\label{ordA}
\end{figure}

Before
presenting the results of simulations for the binary-mass disordered anharmonic
chain it is important to know the value of $\alpha$ for the binary ordered
chain. Let us thus discuss this  first.
This case was
earlier studied in \cite{mai07} where it was found that the temperature
profile showed oscillations with an amplitude that decreased as
$N^{-1/2}$. Let us denote the mass ratio $m_1/m_2=A$. For $A=1$,
the simulations in \cite{mai07} gave strong evidence for an exponent
$\alpha=1/3$. However for the value $A=2.62$ 
a clear convergence could not be obtained. 
Here we will argue that the exponent remains 
unchanged from the $A=1$ value.  
In Fig.~(\ref{ordA}) we present simulation results for the $N$-dependence of the  
current $J$ in the binary-mass ordered chain for different values of
the parameter $A$, all corresponding to the same average mass
$(m_1+m_2)/2=1$. 
Remarkably we  
find that at large enough system sizes the actual values of the
currents for different $A$ tend to converge to the same value as the 
$A=1$ value. Thus clearly  the exponent $\alpha$ remains unchanged for
any value of $A$. However for a  large mass ratio one has to go to large
system sizes to see the true exponent. A similar effect was seen in
Refn.[\onlinecite{grass02}] for the binary hard-particle gas. 
In our simulations we used the velocity-Verlet algorithm with
time steps $dt=0.005$ \cite{AT87}.
For small system sizes we used  $O(10^7)-O(10^8)$ steps for relaxation
and same number of steps for  
averaging, while for larger systems, up to  $O(10^9)$ steps were used. 
In all our simulations we used $T_L=1.25, T_R=0.75$ and $\gamma=1.0$.

\begin{figure}
\vspace{1cm}
\includegraphics[width=3in]{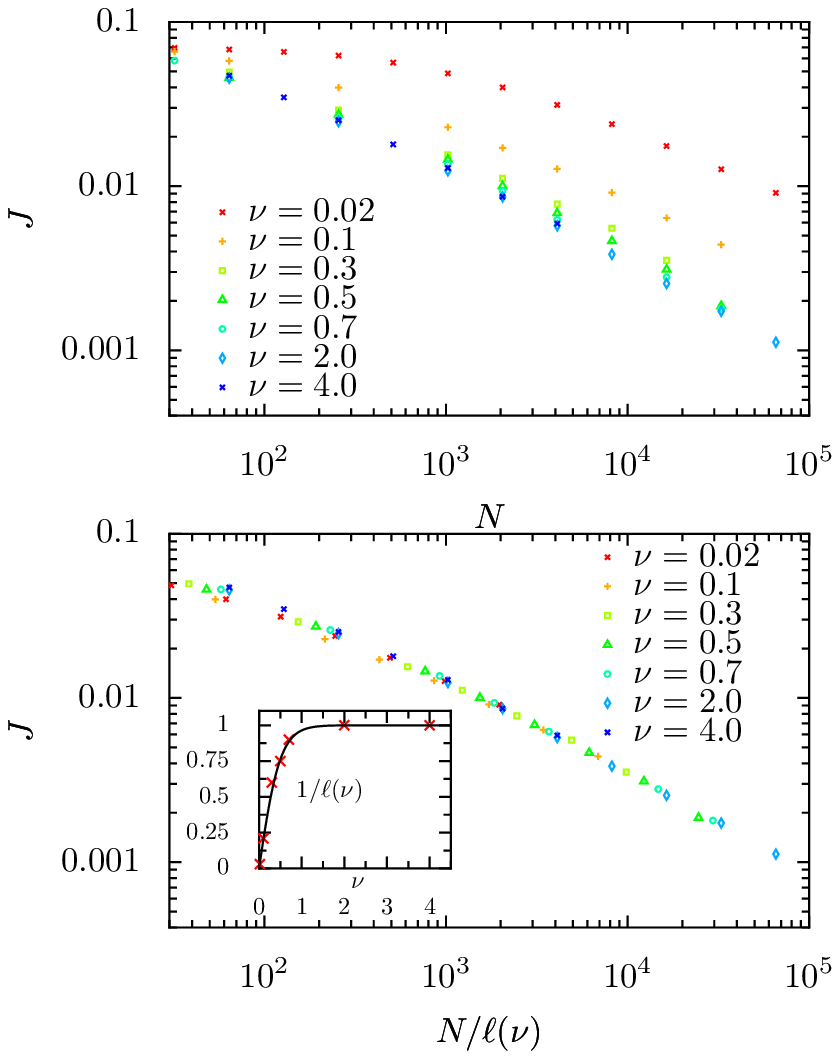}
\caption{Plot of the heat current $J$ versus
system size for the binary mass ordered chain, for different values of
$\nu$ and with the mass ratio $A=1.5$ (a). Fig (b) shows the same data
plotted with a scaled $x$-axis.   
}
\label{ordnu}
\end{figure}

\begin{figure}
\vspace{1cm}
\includegraphics[width=3in]{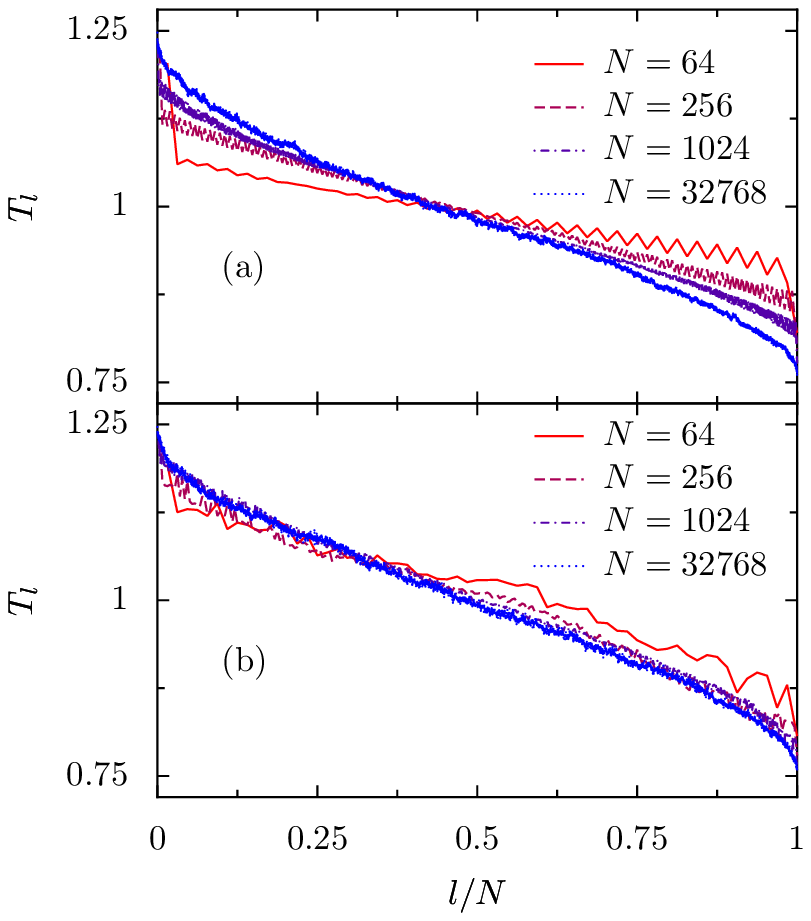}
\caption{Plot of the temperature profiles in the (a) ordered and (b)
  disordered lattices for $\nu=0.1$ and for different system sizes. 
}
\label{temppr}
\end{figure}

For the disordered anharmonic case, we wish to study the cases with
weak and strong anharmonicity and see if there is a transition in the
value of $\alpha$. For small $\nu$ we expect the system's  behaviour
to be close to a harmonic one, and so one would have to go to very large system
sizes to see the effect of anharmonicity and the correct
exponent. It then becomes necessary to try and
understand the data using some sort of a scaling analysis. Let us
first do this for the ordered case. We fix the value of $A=1.5$ and look at the
$N$-dependence of the current for different values of
$\nu$. The results are shown in Fig.~(\ref{ordnu}a). For small system
sizes, we find   
a flat region which is expected since for system sizes smaller than the 
phonon-phonon scattering length scale we expect the system to behave
as a harmonic chain. The scattering length should be larger for smaller $\nu$
and this can be seen in the plot. At large enough
system sizes all curves tend to show the same decay coefficient $\alpha=2/3$.  
To see this clearly we scale the system size by a length factor
$\ell(\nu)$.  Fig.~(\ref{ordnu}b) shows a nice collapse of the data and
the value of the exponent is confirmed. We find empirically that the
$\nu$-dependence of the length parameter is given by $\ell(\nu)= 1/\tanh
(2\nu)$. We note the interesting and somewhat surprising point that 
for any given system size, the value of the current saturates as we
keep increasing $\nu$.  
In Fig.~(\ref{temppr}a) we show the temperature profiles for different
system size 
for $\nu=0.1$. As noted earlier in \cite{mai07} we see the large
oscillations in temperature. An interesting general feature of temperature
profiles in FPU chains is the following. 
A coarse-grained temperature profile obtained by averaging over many
particles would be smooth and monotonic. However the
temperature gradient is non-monotonic and this appears to be true even
for small temperature differences. This implies that it may 
not even be possible to write a phenomenological relation such as $J=
-\kappa_N(T) \nabla T$.

\begin{figure}
\vspace{1cm}
\includegraphics[width=3in]{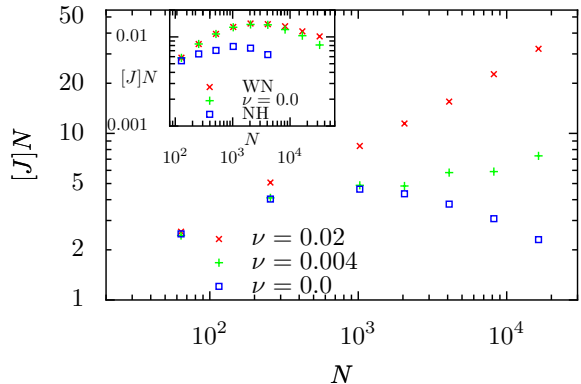}
\caption{Plot of heat current versus system size, for the disordered
anharmonic  chain, for different values of $\nu$. The data in the inset, 
corresponds to parameters $(T_L, T_R )=(0.001,0.0005)$ with 
Gaussian white noise bath for $\nu=1$ (WN) and $\nu=0$, and 
Nose-Hoover bath (NH) for $\nu=1$.}
\label{lowtemp}
\end{figure}

\begin{figure}
\vspace{1cm}
\includegraphics[width=3in]{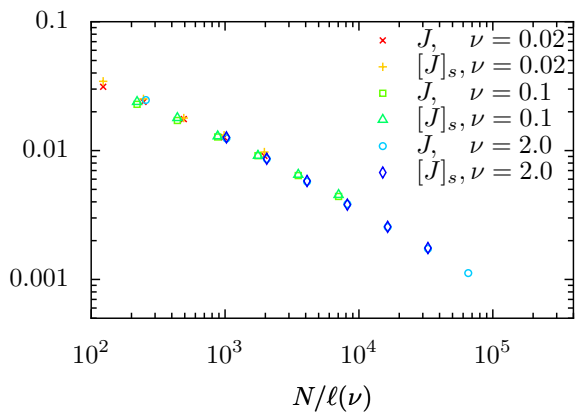}
\caption{Plot of scaled heat current versus scaled system size, for
  the binary-mass ordered ($J$) and disordered ($[J]_s$) anharmonic  chains, for
  different values of $\nu$. 
}
\label{scaling2}
\end{figure}

Finally we now give the results for the disordered anharmonic case. 
We take averages over $50-100$ samples  $N < 1024$, $10$ samples for
$N=1024-16384$, and $2$ samples for $N=32768$ and $65536$. 
In Fig.~(\ref{lowtemp}) we plot the results of simulations for $[J]$ 
for $\nu=0.004$ and $0.02$.
Also we show the result for
$\nu=0.0$. For small values of $\nu$ we see that, at small system
sizes the current value is close to the $\nu=0$ value. As expected
one has to go to large system sizes to see the effect of the weak
anharmonicity. At sufficiently large $N$ we find the same system size
dependence of $[J]$ as obtained for the ordered FPU chain, namely that given by
$\alpha=1/3$. In fact by scaling the current by appropriate factors
we find that the data for the disordered case can be made to collapse
on to the binary-mass ordered case. This is shown in
Fig.~(\ref{scaling2}) (for $\nu=0.02,0.1,2.0$).
Thus our results show that the asymptotic
power law dependence of the current is always dominated by
anharmonicity while disorder only decreases the overall conductance of
a sample. In Fig.~(\ref{temppr}b) we plot the temperature profile for
$\nu=0.1$ and find that the asymptotic profile is similar to the ordered case.
We now throw some light on the reasons which led to the 
erroneous conclusions in \cite{baowenli01}, of a transition in conducting
properties at low temperatures (or equivalently small anharmonicity).
Consider the data for $[J] N$ plotted in Fig.~(\ref{lowtemp}) for
$\nu=0.004$. We see that at around $N\sim 1000-2000$ the data seems to
flatten and if one had just looked at data in this
range, as was done by \cite{baowenli01}, one would conclude that 
Fourier's law is valid. However 
the behaviour changes drastically when one looks at larger system
sizes and one again gets the usual FPU behaviour. To verify that this
is indeed what happens for the particular case studied by Li \etal
we
have repeated simulations with their set of parameters but for much
larger system sizes and the results are shown in the inset of
Fig.~(\ref{lowtemp}). This case corresponds to a much smaller value of
$\nu$ and so it is expected that it will follow the $\nu=0$ curve till
very long length scales and this is clearly seen. However at around $N=16384$ 
we see a tendency
for the curve to turn up and we expect that the same asymptotic
behaviour to eventually show up.  While a transition cannot be ruled
out at even lower temperatures, this seems unlikely. Also, if there is
such a transition, it should probably be to a disordered phase with $[J]\sim
1/N^{3/2}$. 

\section{Temperature dependence of conductivity} 
The scaling property of the current, mentioned
earlier [after Eq.~(\ref{eq:2})], implies that the thermal
conductivity  has the form $\kappa=\kappa(\nu T)$.  For small
anharmonicity ($\nu << 1$), our earlier results imply that at large
system sizes $\kappa \sim (N \nu)^{1/3}$ and from the scaling property
this immediately gives $\kappa \sim T^{1/3}$ at low
temperatures. However at small system sizes  [$ N << \ell (\nu)$], we
expect the system to behave like a harmonic system with $\kappa
\sim T^0$. At high temperatures the conductivity will saturate to a
constant value.

\section{Discussion} Our main conclusion is that there is no change, in the asymptotic
power law dependence of the current on system size, on 
decreasing the temperature in the disordered FPU problem. At low
temperatures one has to go to much larger system sizes to see the true
exponent, whose value ($\alpha=1/3$) is the same as that for the ordered FPU
chain. 
We also find several interesting new results for the binary-mass
ordered FPU chain: (i) the exponent $\alpha$ is independent of the mass
ratio $A$ and is the same as the $A=1$ value, (ii) the data for
different values of $\nu$ can be collapsed by scaling the system size
by a $\nu$-dependent length scale. 
Also we make the interesting and somewhat surprising observation that for a
finite FPU chain, $J \to ~{\rm const}~ \ne 0$ as $\nu \to \infty$. 
Experiments measuring heat conduction in quasi-$1D$
systems such as nanowires, nanotubes \cite{angel98} are now being
done. The effect of isotopic disorder has also been measured
\cite{chang06}. Our prediction is that while disorder will lower the current
in a wire, the system-size dependence of $[J]$ is unaffected.  
Experimentally, the temperature
dependence of the thermal conductivity may be easier to measure
and one can verify if this is unaffected by disorder.

AD thanks Joel Lebowitz for useful discussions. KS was supported by MEXT 
, Grant Number (19740232).

\end{document}